%
%
\input harvmac.tex
\noblackbox
%
%
%
%
\lref\kks{S. Kachru, J. Kumar and E. Silverstein, ``Vacuum Energy
Cancellation in a Non-supersymmetric String,'' hep-th/9807076.}
\lref\ewcc{E. Witten, ``Strong Coupling and the Cosmological Constant,''
Mod. Phys. Lett. {\bf A10} (1995) 2153, hep-th/9506101.}
\lref\ks{S. Kachru and E. Silverstein, ``4d Conformal Field Theories and
Strings on Orbifolds,'' Phys. Rev. Lett. {\bf 80} (1998) hep-th/9802183.}
\lref\ads{J. Maldacena, ``The large N Limit of Superconformal Field Theories
and Supergravity,'' hep-th/9711200.}
\lref\gkpw{S. Gubser, I. Klebanov and A. Polyakov,
``Gauge Theory Correlators
from Noncritical String Theory,'' Phys. Lett. {\bf B428}
(1998) 105, hep-th/9802109 \semi
E. Witten, ``Holography and Anti de Sitter Space,'' hep-th/9802150.}
\lref\vw{C. Vafa and E. Witten, ``Dual String Pairs With $N=1$
and $N=2$ Supersymmetry in Four Dimensions,''  Nucl. Phys.
Proc. Suppl. {\bf 46} (1996) 225,hep-th/9507050.}
\lref\fhsv{S. Ferrara, J. A. Harvey, A. Strominger and C. Vafa,
``Second Quantized Mirror Symmetry,'' Phys. Lett. {\bf B361} (1995) 59,
hep-th/9505162.}
\lref\nsdual{S. Kachru and E. Silverstein, ``$N=1$ Dual Pairs and
Gaugino Condensation,''  Nucl. Phys. {\bf B463}
(1996) 369, hep-th/9511228; O. Bergman and M. R. Gaberdiel, ``A
Nonsupersymmetric Open String Theory and S Duality,'' Nucl.
Phys. {\bf B499} (1997) 183, hep-th/9701137;
J. Blum and K. Dienes, ``Strong/Weak Coupling Duality and Relations for
Nonsupersymmetric String Theories,'' Nucl. Phys. {\bf B516} (1998) 83,
hep-th/9707160, and ``Duality Without Supersymmetry: The Case of the
$SO(16) \times SO(16)$ String, '' Phys. Lett. {\bf B414} (1997) 260,
hep-th/9707148.}
\lref\hullt{C. M. Hull and P. K. Townsend, ``Unity of String Dualities,''
Nucl. Phys. {\bf B438} (1995) 109, hep-th/9410167.}
\lref\wittdyn{E. Witten, ``String Theory Dynamics in Various Dimensions,''
Nucl. Phys. {\bf B443} (1995) 85, hep-th/9503124.}
\lref\hs{J. A. Harvey and A. Strominger,
 ``The Heterotic String is a Soliton,'' Nucl. Phys. {\bf B458} (1996) 456,
hep-th/9504047.}
\lref\senstr{A. Sen, ``String-String Duality Conjecture in Six Dimensions
and Charged Solitonic Strings,'' Nucl. Phys. {\bf B450} (1995) 103,
hep-th/9504027.}
\lref\ds{M. Dine and E. Silverstein, ``New M-theory Backgrounds with
Frozen Moduli,'' hep-th/9712166.}
\lref\hotstr{R. Rohm, ``Spontaneous Supersymmetry Breaking in
Supersymmetric String Theories,'' Nucl. Phys. {\bf B237} (1984)
553 \semi J. Polchinski, ``Evaluation of the One Loop String
Path Integral,'' Commun. Math. Phys. {\bf 104}
(1986) 37 \semi B. McClain and B.D.B. Roth, ``Modular Invariance for
Interacting Bosonic Strings at Finite Temperature,'' Commun. Math.
Phys. {\bf 111} (1987) 539 \semi
K. H. O'Brien and C.-I. Tan, Phys. Rev. {\bf D36} (1987) 1184.}
\lref\itt{H. Itoyama and T. Taylor, Phys. Lett. {\bf B186} (1987) 129.}
\lref\kutsei{D. Kutasov and N. Seiberg, ``Number of Degrees of Freedom,
Density of States and Tachyons in String Theory and CFT,'' Nucl. Phys.
{\bf B358} (1991) 600.}
\lref\nsmod{V. P. Nair, A. Shapere, A. Strominger and F. Wilczek,
``Compactification of the Twisted Heterotic String,'' Nucl. Phys.
{\bf B287} (1987) 402 \semi
P. Ginsparg and C. Vafa, ``Toroidal Compactification of
Nonsupersymmetric Heterotic Strings,'' Nucl. Phys. {\bf B289} (1987) 414.}
%
%

\Title{\vbox{\baselineskip12pt
\hbox{EFI-98-31}
\hbox{hep-th/9807213}
}}
{\vbox{\centerline{String Duality}
\centerline{and}
\centerline{Non-supersymmetric Strings}}}

\centerline{Jeffrey A. Harvey}
\centerline{Enrico Fermi
Institute and Department of Physics}
\centerline{University of
Chicago, 5640 Ellis Avenue}
\centerline{Chicago, IL 60637}

\bigskip
\centerline{\bf Abstract}

In recent work Kachru, Kumar and Silverstein introduced a special class of
non-supersymmetric type II string theories in which the cosmological
constant vanishes at the first two orders of perturbation theory. Heuristic
arguments suggest the cosmological constant may vanish in these theories
to all orders in perturbation theory leading to a flat potential for
the dilaton. A slight variant of their
model can be described in terms of a dual heterotic theory.
The dual theory has a non-zero cosmological constant
which is non-perturbative in the coupling of the original type II theory.
The dual theory also predicts a mismatch between Bose and Fermi degrees of
freedom in the non-perturbative D-brane spectrum of the type II theory.

\Date{July 28, 1998}

%
\newsec{Introduction}
In spite of great recent progress in the understanding of string
theory and quantum gravity the smallness of the cosmological constant
remains a great mystery. At present the only obvious explanation for
a vanishing cosmological constant is  unbroken
supersymmetry. Since supersymmetry is broken if it is realized in
nature the puzzle becomes why the cosmological constant is so much
smaller than the scale set by the supersymmetry breaking scale, that
is why $\Lambda << (TeV)^4$.

Motivated by the AdS/CFT correspondence \refs{ \ads, \gkpw}
Kachru and Silverstein
have suggested \ks\  that the cosmological constant $\Lambda$
might vanish in
certain special non-supersymmetric string theories and together with
Kumar have constructed a candidate such theory \kks. The one-loop
contribution to $\Lambda$ vanishes trivially in the
model of \kks\ due to equality between the number of boson and fermion
mass states at each level (in spite of the fact that the model
is not supersymmetric).  What is more surprising is the claim that
the two-loop and perhaps higher loop contributions also vanish since
without supersymmetry one would expect the non-supersymmetric
interactions to spoil the
cancellation at some order in perturbation theory.
Unfortunately the intricacies of higher-loop calculations in fermionic
string theory make a direct analysis of this claim  difficult.
In addition, if the perturbative contribution to  $\Lambda$ does indeed
vanish it will be important to investigate non-perturbative contributions
and their dependence on the string coupling constant.

In this paper an indirect approach to this problem is taken using
string duality. The heuristic arguments of \refs{\ks, \kks}
suggest that the
model of \kks\ has a vanishing dilaton potential and therefore
exists at all values of the coupling. If this were the case we
should be able to use string duality to study the strongly coupled
limit of this theory. As it turns out, we will find evidence that
the cosmological constant is non-zero so that there is a potential for the
dilaton. Because of this the theory probably does not have a stable
vacuum at strong coupling.

In spite of this one might hope to study some features of the
theory using string duality.  For example the theory presumably
has cosmological solutions in which the dilaton evolves in time
and one could try to use string duality to explore the dilaton
potential in different regimes. In the models at hand the dilaton
potential vanishes in a limit where supersymmetry is restored
and one can try to use duality to study the theory near this
limit. Of course the use of string duality
is on much less firmer
ground in theories without spacetime supersymmetry because one loses
the BPS states and non-renormalization theorems which provide the
most direct evidence for string duality. Nonetheless there are some
rough indications that duality can be applied in this context \nsdual.
In addition the string-string duality that will be applied here is
well understood in the supersymmetric
case \refs{\hullt,\wittdyn,\hs,\senstr}
and the adiabatic argument
of \vw\ can be applied to the non-supersymmetric dual pairs which
are constructed.  These facts plus the nature of the results found
here give some indication that string duality can be successfully
used to study models of the type discussed in \kks.

\newsec{A Non-supersymmetric String}

Following \kks\ consider type IIA string theory compactified on a six-torus
$T^6$. We take the first four components to be a square torus at
the self-dual radius $T^4 = (S^1_R)^4$ with $R= 1/\sqrt{2}$ (the string
scale has been set to one).
The last two components we take to be the
product of two circles of radius $R_5$ and $R_6$. We now consider
an asymmetric orbifold of this theory. An element of the space group
of the orbifold will be denoted by
\eqn\onea{[(\theta_L),(\theta_R),(v_L),(v_R), \Theta]}
where $\theta_{L,R}$ are rotations by elements of $SO(6)$ acting
on the left or right-moving degrees of freedom and $v_{L,R}$ are
shifts acting on the left or right-moving bosons. In the examples
considered here $\theta_{L,R}$ are order two and will be denoted by
listing their eigenvalues. $\Theta$ will
either be the identity or $(-1)^{F_{L,R}}$ which are rotations by
$2 \pi$ acting on the left or right-moving degrees of
freedom and are therefore $+1$ on
left or right-moving bosonic states and $-1$ on left
or right-moving fermion states.

The
asymmetric orbifold is generated by the following two elements f and g:
\eqn\oneb{\eqalign{f & =[(-1^4,1^2),(1^6),
(0^4,v_L^5,v_L^6),(s^4,v_R^5,v_R^6)
 (-1)^{F_R}] \cr
 g & = [(1^6),(-1^4,1^2),
 (s^4,w_L^5,w_L^6),(0^4,w_R^5,w_R^6)
  (-1)^{F_L}] \cr }}
Here a power indicates repeated entries (except on $w$ or $v$) and
$s$ is a shift by $R/2=1/2 \sqrt{2}$ and modular invariance
(level matching) requires that
\eqn\onec{ (v_L^5)^2+(v_L^6)^2 - (v_R^5)^2 - (v_R^6)^2 =
(w_L^5)^2+(w_L^6)^2 - (w_R^5)^2 - (w_R^6)^2 =1/2 }
The model considered in \kks\ arises by setting $R_5=R_6= 1/\sqrt{2}$ and
taking  $v_L^5=1/\sqrt{2},v_R^5=0,v_L^6=v_R^6=1/2 \sqrt{2}$ and
$w_L^5=w_R^5 = 1/2\sqrt{2}, w_L^6=0,w_R^6= 1/\sqrt{2}$.

The shifts can be described more transparently in the notation of
\vw. Points in the Narain lattice $\Gamma^{1,1}$ are described by
pairs of integers $(m,n)$ with $m$ labeling the momentum and $n$ the
winding. There are three choices of shift vector $A$ with $2A \in
\Gamma^{1,1}$ modulo vectors in $\Gamma^{1,1}$:
$A_1=(1/2,0)$, $A_2=(1/2,1/2)$ and $A_3=(0,1/2)$. The  model of \kks\
has shifts by $A_1$ in the first four components of $T^6$ for
both $f$ and $g$, a shift in the fifth component by $A_2$ in
$f$ and by $A_1$ in $g$, and a shift in the sixth component by
$A_1$ in $f$ and by $A_2$ in $g$.

Besides
satisfying level-matching, the shifts in the model of \kks\ are chosen
to ensure that there are no massless states in sectors twisted by
$f$,$g$ or $fg$. $f$ and $g$ project out the gravitinos coming from
the right
and left-movers and without shifts the gravitinos would come back
in the sectors twisted by $f$ or $g$. The sector twisted by $fg$ on the
other hand does not lead to gravitinos even without shifts.

Now consider a slight variation of the previous model in which
the shift $w_{L,R}^5$ is exchanged with $w_{L,R}^6$ and denote the
resulting generators by $f'$ (which is the same as $f$) and $g'$.
In addition since the
radii of the last two components of $T^6$ are not fixed by
the asymmetric orbifold we should allow arbitrary
radii $R_5$ and $R_6$ in these components.
Furthermore, after this modification, the last component of $T^6$ is
irrelevant to the construction, the shift in this circle is by $A_1$
in both $f$ and $g$ and is not needed for modular invariance or
to ensure that there are no massless states in the sectors twisted
by $f$ or $g$. For simplicity we might as well take $R_6 \rightarrow
\infty$ and consider the resulting five-dimensional variant of
the previous model. We will also denote $R_5$ by $R'$ from here on.

This gives a model which
like the model of \kks\ is not supersymmetric and which has equal numbers
of bosons and fermions at each mass level at string tree level. Most
of the higher loop analysis of \kks\  also appears to
be unchanged by this modification although this has not
been investigated in detail. Also, in \kks\ a heuristic argument
for the vanishing of $\Lambda$ was given using the AdS/CFT correspondence.
This argument involves the existence of Reissner-Nordstrom black holes
with $AdS_2$ near horizon geometry. This in turn requires the vanishing
of couplings of the form $\int \phi F^2$ with $\phi$ a modulus and $F$
the gauge field of the $U(1)$ gauge symmetry under which the black hole
is charged. In the model of \kks\ such black holes can be constructed
as linear combinations of $D$-brane states  which are
invariant under the orbifold group \ds. These give rise to RN black holes
which carry charge under a $U(1)$ which arises in the untwisted, RR sector
of the orbifold. The modification made here leads to new moduli in the
$fg$ twisted sector of the orbifold. However these new moduli cannot
have couplings of the form $\phi F^2$ since $\phi$ arises in a twisted
sector while $F$ comes from the untwisted sector. Thus the heuristic
argument of \kks\ also goes through with this modification.

Finally, it is important to note that $f$ and $g$ do not commute as elements
of the space group $S$ either in the original model
or in the modification considered here. However in constructing a
string orbifold
we can first mod out the theory in $R^d$ by the lattice $\Lambda$,
which is the normal subgroup of $S$
consisting of all pure translations
in the space group. We can  then mod out the resulting
theory by the point group $\bar P = S/\Lambda$.
In the one-loop string path integral it is necessary to sum over
twist structures on the world-sheet torus by commuting pairs elements of
the orbifold group. Thus from the space group point of view there
are no sectors with boundary conditions $(f,g)$. On the other hand,
$f$ and $g$ do commute as elements of $\bar P$ so that if we first mod out
by $\Lambda$ and then mod out by $\bar P$
 we do expect to have sectors with boundary
conditions $(f,g)$. These
two points of view are reconciled by the fact that
the contribution from the sector $(f,g)$
vanishes due to the fact that the trace of  $f$ is zero in the Hilbert
space twisted by $g$ and vice versa. The fact that $f$ and $g$ commute
in the point group also means that we can first consider the theory
twisted by the product $fg$ and then mod out this theory by $f$.

\newsec{Analysis of the Modified Model}

Now consider the modified  model constructed by first modding out
by the product $ f'g'$ and the modding out by $f'$.
The product $f'g'$ is given
by
\eqn\twoa{f'g' = [(-1^4,1),(-1^4,1),
(0^5),(0^5),(-1)^{F_L+F_R}]}
The shifts in the product $f'g'$ are shifts by elements of the lattice
and can thus be taken to be zero in the point group. The $(-1)^4$
action is a twist by an element in the center of one of the $SU(2)$
factors in the decomposition $SO(4)=SU(2) \times SU(2)$ of the rotation
acting on the first four coordinates. Furthermore,
the factor of $(-1)^{F_L+F_R}$ can be dropped since it is simply
a $2 \pi$ rotation on both left and right coordinates and so can
be absorbed by a choice of the $(-1)^4$ action.

Thus the theory twisted by $f'g'$ is just IIA theory on
$T^4/Z^2 \times S^1$
which is an orbifold limit of IIA theory on $K3 \times S^1$. The massless
spectrum of this theory is just the naive dimensional reduction on the
$S^1$ of IIA
theory on $T^4/Z_2$. In terms of representations of $D=6$, $(1,1)$
supersymmetry the massless spectrum on $T^4/Z_2$
 consists of the graviton supermultiplet
$G_{(1,1)}$ and four copies of the matter multiplet $\Phi_{(1,1)}$ from
the untwisted sector and sixteen copies of the matter multiplet
$\Phi_{1,1}$ coming from the twisted sector. In terms of the massless
little group $SO(4)=SU(2) \times SU(2)$ these representations decompose as
\eqn\twob{\eqalign{\Phi_{(1,1)} = & [(2,2)+2(2,1)+2(1,2)+4(1,1)] \cr
G_{(1,1)}  = & [(3,3)+ 2(3,2)+2(2,3)+(3,1)+(1,3)+4(2,2) \cr
 & +
2(1,2)+2(2,1)+(1,1)] \cr  = & [(2,2)] \times \Phi_{(1,1)} \cr }}

We now twist this theory by $f'$ which acts as a twist by $-1$ on each
of the four left-moving coordinates of $T^4$, as a shift by $A_1$ on each
of the four components of $T^4$, as a shift by $A_2$ on the $S^1$ and
as $(-1)^{F_R}$ on the right-moving degrees of freedom. In the untwisted
sector this implies that $f'$ has eigenvalue $\pm 1$ on
the bosons (fermions) in $G_{(1,1)}$ and eigenvalue $\pm 1$ on the fermions
(bosons) in $4 \Phi_{(1,1)}$ coming from the untwisted sector. Acting
on the states in the sector twisted by $f'g'$, the shifts in $f'$ permute
the 16 fixed points so that the trace of $f'$ vanishes. Taking this
and the $(-1)^4_L (-1)^{(F_R)}$ action into account one sees that
$f'$ acts as $\pm 1$ on the bosons (fermions) for $8$
linear combinations of the $ 16 \Phi_{(1,1)}$
and as $\mp 1$ on the bosons (fermions) for the other $8$ orthogonal
linear combinations. Thus projecting onto $f'$ invariant states leads to
equal numbers of massless (and massive) bosons and fermions in the
sector twisted by $f'g'$.

Because of the asymmetric shifts in $f'$ and $g'$ there are no massless
states in the sectors twisted by $f'$ and by $g'$ separately. As in
the model of \kks\ the massive states are Bose-Fermi degenerate
in these sectors as well.

\newsec{The Heterotic Dual}

We have seen that the type IIA theory twisted by $f'g'$ is an orbifold
limit of type IIA on $K3 \times S^1$ and this is known to be dual
to heterotic string on $T^4 \times S^1$ \refs{\hullt,\wittdyn,\hs,\senstr}.
It will be useful in what follows to recall some facts from
the detailed discussion in \wittdyn. Consider the heterotic string
compactified on a fixed
$T^4$ and a circle $S^1_R$ of radius $R$ and with coupling $\lambda$.
This is dual to a IIA string with coupling $\lambda'$ on a fixed $K3$
and a circle of radius $R'$ or after T-duality to a IIB theory with
coupling $\lambda''$ on a fixed $K3$ and a circle of radius $R''$.
The relation between the parameters is
\eqn\threea{\eqalign{\lambda' & = 1/\lambda,  \qquad R' = R/\lambda \cr
                     \lambda'' & = 1/R, \qquad R'' = \lambda/R \cr }}
The mapping between charged states in the heterotic and IIA theories
is such that states with momentum on the $S^1$ in the heterotic string
map to momentum states in IIA, winding heterotic states map to wrapped
fivebranes in IIA, wrapped heterotic fivebranes map to winding states
in type IIA and perturbative states charged under the original
ten-dimensional
gauge fields of the heterotic string map to D0-brane states in the
five-dimensional IIA
theory (some of which arise from wrapping D2 and D4-branes on the $K3$).
The mapping of states in the IIB description follows from the standard
action of T-duality on branes.

Now given a pair of dual theories it is often possible to construct
further dual pairs by orbifolding \fhsv. This procedure is most reliable
when the orbifold symmetry acts freely on an $S^1$ so that the adiabatic
argument of \vw\ can be used. This is true in the case at hand. We mod
out the type II theory on $T^4/Z_2 \times S^1$ by the action of $f'$ in
order to obtain a non-supersymmetric string. Since $f'$ acts as a shift
by $A_2$ on the $S^1$ we can apply the adiabatic argument.

In order to construct the heterotic dual we need to know the image $f_H$
of $f'$ under duality. Given the action of $f'$ on the $U(1)^{26}$ gauge
bosons of the type II theory on $T^4/Z_2 \times S^1$ it is easy to see
that up to shifts $f_H$ must have twelve $+1$ eigenvalues and twelve
$-1$ eigenvalues when acting
on the left-moving degrees of freedom of the heterotic string and act
as $(-1)^{F_R}$ on the right-moving degrees of freedom.  On the 
type II side $f'$ exchanges the $16$ fixed points in the $f'g'$
twisted sector with each other. On the heterotic side this maps
to an action of $f_H$ which exchanges two $E_8$ lattices on the
left. The remaining four $-1$ eigenvalues must then come
from a $-1^4$ action on four left-moving coordinates of a
$\Gamma^{4,4}$ lattice. String-string
duality does not give a unique prescription for the shifts, in part
because perturbative shifts on the heterotic side map to RR gauge
transformations on the type II side which are not visible in perturbation
theory \fhsv. The shifts on the heterotic side must then be determined by
demanding level matching. In this case $f_H$ acts as $-1$ on 12 left-moving
bosons and this raises the vacuum energy by $12/16=3/4$. This can be
compensated by a shift by $A_1$ in each of the four components of
the $\Gamma^{4,4}$ lattice \foot{If we take the $\Gamma^{4,4}$ lattice
to be at the $SU(2)$ self-dual point in all four coordinates then $f_H$ is
order four. Its square is a translation and orbifolding by this
translation generates a $\Gamma^{4,4}$ lattice at the $SO(8)$
enhanced symmetry point. Acting on this new lattice $f_H$ is then
order two.}.

Finally, modular invariance is consistent with a shift in the
remaining $\Gamma^{1,1}$ lattice by either $A_1$ or $A_3$. We
choose the shift to be $A_3$ so that supersymmetry is restored
at large radius. The T-dual theory would have a shift by $A_1$
and have supersymmetry restored at small $R$. 

To summarize, we consider a point in the Narain moduli space of
the heterotic string on $T^4 \times S^1$ where the Narain lattice can
be decomposed as \foot{This is not the same point in moduli space as
the type II dual and is chosen to simplify the presentation.}
\eqn\threeb{\Gamma^{21,5} = \Gamma^{8,0} \oplus \Gamma^{8,0} \oplus
\Gamma^{4,4}(D_4) \oplus \Gamma^{1,1}(R)}
Here $\Gamma^{8,0}$ is the $E_8$ lattice, $\Gamma^{4,4}(D_4)$
is the $(4,4)$ Narain lattice at the $D_4=SO(8)$ enhanced symmetry
point and $\Gamma^{1,1}$
is the Narain lattice for compactification on a $S^1$ of radius $R$.
Then in terms of this decomposition $f_H$ acts as an interchange of the
two $\Gamma^{8,0}$ factors,
as $-1^4$ on the left-moving degrees of freedom and a shift
by $A_1^4$ on the right-moving degrees of freedom of the third
component and as
a shift by $A_3$ on the fourth component in addition to the
$(-1)^{F_R}$ action
on the right-movers.

Twisting the heterotic theory by $f_H$ breaks all the supersymmetry
since it projects out the gravitino coming from the right-movers. As
in the type II construction, supersymmetry is restored in the large
radius limit.
The massless spectrum
in the untwisted sector
has equal numbers of bosons and fermions. To see this note
that since $f_H$ is $\pm 1$ on
right moving bosons (fermions) and has an equal number of $\pm 1$
eigenvalues
on the left the massless states with one left-moving
oscillator state excited come in Bose-Fermi pairs. Similarly, $f_H$
has 252 eigenvalues $+1$ and $252$ eigenvalues  $-1$
acting on $P_L^2=2$ states
in the
$\Gamma^{8,0} \times \Gamma^{8,0} \times \Gamma^{4,4}(D_4)$ lattice so
these states also contribute equal numbers of bosons and fermions.
Note that there is a low-energy non-abelian $E_8$ gauge theory
with the same field content as that of $N=4$ supersymmetric gauge theory. 
In the heterotic theory there are states with momentum $m/R$
which become  massless as $R \rightarrow \infty$. From the previous
argument it is clear these states are also Bose-Fermi degenerate.
Since these states map to perturbative momentum states in the type
II theory which are Bose-Fermi degenerate this is required for the
duality to act correctly.

Now let us study the spectrum of states of this asymmetric heterotic
orbifold which are massive at large $R$.
First consider the twisted sector.
In the twisted sector all states are massive at a
generic radius $R$ as a result of the shift by $A_3$.
Since there are $12$ anti-periodic
bosons on the left the left-moving vacuum energy is
$E_L = -1/4 + ({A_3}_L)^2/2  $. The right-moving fields are untwisted
but have a shift in the $\Gamma^{4,4}$ which raises the vacuum energy
by $1/4$ 
so the right-moving vacuum energy is $E_R =-1/4+ ({A_3}_R)^2/2$
in the Neveu-Schwarz sector and $E_R = 1/4+({A_1}_R)^2/2 $ in the
Ramond sector. Because of the $(-1)^{F_R}$ action we must also change
the GSO projection so that the NS vacuum now survives the GSO projection.
We thus see that this theory has a tachyon for  $R<1/2 \sqrt{2}$.
In the Ramond sector there are of
course no tachyons, so there is a mismatch between bosons and fermions in
the twisted sector. This mismatch clearly continues to exist at large
values of $R$ where there is no tachyon.

There is also a mismatch between bosons and fermions in the untwisted sector
of the orbifold for states which are massive at large $R$.
This can be seen just be writing down the first few massive
states or can be summarized by the one-loop partition function in the
untwisted sector. This is given by $(Z_{1,1} + Z_{f_H,1})/2$ where $Z_{a,b}$
is the one-loop torus amplitude with boundary condition twisted by $a$
in the time direction and by $b$ in the space direction and the usual
sum over fermion spin structures has been suppressed. $Z_{1,1}=0$
since supersymmetry is only broken by the $f_H$ projection. On the
other hand
\eqn\threec{\eqalign{Z_{f_H,1}(\tau) = &
{ \Theta_{2E_8}(q) \over \eta(q^2)^{12} \eta(\bar q)^8 }
\theta_4^4(\bar q^2)
\left( \sum_{p\in \Gamma^{1,1}} q^{p_L^2/2} {\bar q}^{p_R^2/2}
e^{2 \pi i p \cdot A_1} \right) \cr
 & \times \left( { \theta_3^4(\bar q) \over {\bar \eta}^4 } -
{ \theta_4^4(\bar q) \over {\bar \eta}^4 } +
{ \theta_2^4(\bar q) \over {\bar \eta}^4 } \right) \cr }}
where $q=e^{2 \pi i \tau}$ with $\tau$ the modular parameter of the
world-sheet torus. $\Theta_{2 E_8}$ is the theta function for the
$E_8$ lattice with norm rescaled by $2$,
and standard notation
has been used for the other
theta functions and for the Dedekind $\eta$ function.
Writing out the first few terms in the $q, \bar q$ expansion of $Z_{f_H,1}$
shows a mismatch between bosons and fermions at massive levels.

Massive charged states in the heterotic theory map to charged
non-perturbative wrapped brane states in the type II theory. Although
there is no BPS formula protecting the mass, the lightest state
of a given charge must be stable even without supersymmetry and so
we should be able to compare these states in the type II and heterotic
descriptions. Thus the mismatch in the heterotic theory in the untwisted
sector predicts a mismatch in the D0-brane spectrum of the type II theory
and the mismatch in the twisted sector implies a mismatch in the wrapped
fivebrane states of the IIB theory.

It is also interesting to compute the cosmological constant in the
heterotic theory in order to compare with the perhaps vanishing
perturbative contribution in the type II theory. The cosmological constant
is proportional to the vacuum amplitude
\eqn\threed{\Lambda \sim  \int_{\cal F} {d^2 \tau \over \tau_2}
 (\alpha' \tau_2)^{-5/2}
\left(Z_{1,1}(\tau) + Z_{f_H,1}(\tau) +
Z_{1,f_H}(\tau) +
Z_{f_H,f_H}(\tau) \right) }
where ${\cal F}$ is the fundamental domain for the modular group,
$|\tau|>1, |\tau_1|<1/2$.
As discussed above $Z_{1,1}=0$ by supersymmetry. The latter two terms
in \threed\ can be determined from \threec\ using the modular
transformations $\tau \rightarrow -1/\tau$ and $\tau \rightarrow \tau+1$.

The analysis of the cosmological constant in this theory is mathematically
very similar to the analysis of the free energy of superstrings
at temperature $T \sim 1/R$ \hotstr.
At small $R$ there is a divergence
in $\Lambda$ coming from the tachyon.
We can however examine the large $R$ behavior of $\Lambda$
and compare to  type II theory using the duality relations \threea.

Using the fact that the three terms contributing to $\Lambda$ are
related by the modular transformations $\tau \rightarrow -1/\tau$
and $\tau \rightarrow \tau+1$ we can write \threed\ as an integral
of $Z_{f_H,1}$ over the fundamental domain of the $\Gamma_0(2)$
subgroup of the modular group $\Gamma$. Denoting this by ${\cal F}_2$
we then have
\eqn\threey{ \Lambda \sim
\int_{{\cal F}_2} {d^2 \tau \over \tau_2^{7/2} } F(q,\bar q)
\sum_{m,n} (-1)^{m} e^{2 \pi i \tau_1 mn }
e^{- \pi \tau_2 (m^2/2R^2 + 2n^2 R^2 ) }
}
where $F(q,\bar q)$ stands for the terms in \threec\ other than the
sum over the $\Gamma^{1,1}$ lattice. From the previous comments we
know that $F$ takes the form
\eqn\threeg{F(q,\bar q) = \sum_{i,j} d(i,j) q^i \bar q^j =
16 \left( q^{-1} + 252 q + \cdots \right) \left( 1 + 8 \bar q + \cdots
\right) }
Note the absence of a $q^0$ term in the $q$ expansion of $F$ which
indicates Bose-Fermi degeneracy for massless states.
In this form the behavior of the integral at large $R$ is not
evident because many terms contribute at large $R$. However we can use
Poisson resummation on $m$ to rewrite the double sum in
\threey\ in the form
\eqn\threeh{\eqalign{\sum_{m,n} & (-1)^{m} e^{2 \pi i \tau_1 mn }
e^{-\pi \tau_2 (m^2/2 R^2 + 2 n^2 R^2) } \cr
& = R \sqrt{2/\tau_2}
\sum_{n,m'}
e^{ -2 \pi R^2 [\tau_2 n^2 + (m'-1/2-n \tau_1)^2/\tau_2]  } \cr}
}
We then need to evaluate the integral
\eqn\threex{\int_{{\cal F}_2} {d^2 \tau \over \tau_2^{7/2}}
R \sqrt{2/\tau_2} \sum_{i,j}\sum_{n,m'}  d(i,j) q^i \bar q^j
e^{-2 \pi R^2 [\tau_2 n^2 + (m'-1/2-n \tau_1)^2/\tau_2] } }

The $n=0$ term in \threex\
can be evaluated by saddle point approximation.
The saddle point is
at large $\tau_2 \sim |m'-1/2|R/M_i$
where the integral over $\tau_1$ restricts to
states with mass $M_i \sim \sqrt{i}=\sqrt{j}$. 
Thus the contribution from states of
mass $M_i$ is of order $e^{-M_i R}$.
Heuristically this can be thought
of as the contribution from a world-line instanton where a particle
of mass $M_i$ has its world-line wrapped around the circle of radius
$R$ in spacetime. This represents the contribution of a single state
and one might worry that the exponential degeneracy of states in
string theory might overwhelm the $e^{-R}$ suppression. This is
equivalent to finding a tachyon in the spectrum and so does not
happen for sufficiently large $R$.
For $n \ne 0$ the saddle point is not necessarily
in ${\cal F}_2$ and in order to evaluate the integral it is necessary
to use the unfolding technique of \hotstr\ to rewrite \threex\
as an integral over the strip $\tau_2 >0, |\tau_1 |<1/2$. This
again leads to an asymptotic behavior
$\Lambda \sim R^{-5/2}e^{-R}$ at large $R$ \foot{I thank
D. Kutasov and G. Moore for help in the evaluation of this
integral}.

Because of the Bose-Fermi degeneracy among massless states and states
which become massless as $R \rightarrow \infty$ the usual power law
behavior of $\Lambda$ at large $R$ cancels out and we are left only
with contributions decreasing as $R^{-5/2} e^{-R}$.
This exponential suppression of $\Lambda$
at large $R$ in models with equal numbers of massless fermions
and bosons and supersymmetry broken by twisted boundary
conditions was noted previously in \itt.

To compare to the type II theory we can take large $R$ in the
heterotic theory. Since this takes us to a six-dimensional
theory we should hold the six-dimensional heterotic coupling
small and fixed. In type IIB language this corresponds to very
small $R''$ so that it is  more natural to compare to the IIA
theory with radius $R'$ which is becoming large. The $e^{-R}$
behavior we have found in the heterotic theory then  becomes
$e^{-R'/\lambda'}$ in IIA variables. This
result is compatible with the vanishing of $\Lambda$ to all orders
in perturbation theory conjectured in \kks.
It is natural to interpret an $e^{-R'/\lambda'}$ effect
in the IIA theory as coming from a world-line instanton where
a D0-brane world-line wraps the $S^1$. It would be interesting to
investigate such effects directly in the type II theory.

\newsec{Comments and Conclusions}

By considering a slight variant of the model considered in \kks\ it
is possible to construct a heterotic dual theory to a non-supersymmetric
string with many if not all of the features of the
model of \kks. The heterotic
dual has a mismatch between Bose and Fermi degrees of freedom at
the massive level. If duality is a reliable guide to the physics of
the type II non-supersymmetric theory then this mismatch implies
a similar mismatch in the type II theory for non-perturbative states
which arise either as D0-branes in $D=5$ (including wrapped $D4$ branes
and D-brane states arising from the twisted sector of the asymmetric
orbifold) or from wrapped fivebranes. It is clearly of some interest
to develop brane technology on asymmetric orbifolds in order to test
whether this is indeed true.  Conversely, if a mismatch is found among
these states in the type II theory it will provide evidence for
the reliability of duality in the absence of spacetime supersymmetry.

The model of \kks\ was motivated by the AdS/CFT correspondence applied
to Reissner-Nordstrom black holes with $AdS_2$ near horizon geometry.
It would also be interesting to see whether a more detailed analysis
of the correspondence in this situation sheds some light on the presence
of non-perturbative corrections. Such an analysis might also suggest
models in which these contributions are eliminated or where the dilaton
is stabilized so that the contributions are exponentially small.

In addition to the radius $R$ the heterotic theory also has moduli
obtained by putting equal Wilson lines in the two $E_8$ factors or
by turning on equal constant metric and antisymmetric tensor
fields $g_{i5}=B_{i5}$ with $i=1, \cdots 4$ labeling the
$\Gamma^{4,4}$ directions and $5$ labeling the $\Gamma^{1,1}$
direction. It
would be interesting to explore $\Lambda$ and its stationary points
as a function of these moduli as in \nsmod.

\bigskip
\centerline{\bf Acknowledgments}\nobreak
\bigskip

I would like to thank A. Dabholkar, M. Green,
D. Kutasov and E. Martinec for helpful discussions.
I am
particularly indebted to S. Kachru, G. Moore and E. Silverstein
for detailed discussions of several important points.
I am grateful to
the participants of the Amsterdam Summer Workshop for useful
discussions and questions during a presentation of a preliminary
version of this work.
In addition I thank the organizers of the SUSY '98 meeting
at Oxford, the organizers of the
Amsterdam Summer Workshop on String Theory,
and the proprietors of de jaren for
hospitality while portions of this work were completed. This work
was supported in part by
NSF Grant No.~PHY 9600697.

\listrefs

\end